# Exploring Physics of Ferroelectric Domain Walls in Real Time: Deep Learning Enabled Scanning Probe Microscopy

Yongtao Liu,[1] Kyle P. Kelley,[1] Hiroshi Funakubo,[2]
Sergei V. Kalinin,[3,a] and Maxim Ziatdinov [1, 4, b]

[1] Center for Nanophase Materials Sciences, Oak Ridge National Laboratory, Oak Ridge, TN 37923, USA

[2] Department of Material Science and Engineering, Tokyo Institute of Technology, Yokohama 226-8502, Japan

[3] Department of Materials Science and Engineering, University of Tennessee, Knoxville, TN 37996, USA

[4] Computational Sciences and Engineering Division, Oak Ridge National Laboratory, Oak Ridge, TN 37923, USA

The functionality of ferroelastic domain walls in ferroelectric materials is explored in real-time via the *in-situ* implementation of computer vision algorithms in scanning probe microscopy (SPM) experiment. The robust deep convolutional neural network (DCNN) is implemented based on a deep residual learning framework (*Res*) and holistically-nested edge detection (*Hed*), and ensembled to minimize the out-of-distribution drift effects. The DCNN is implemented for real-time operations on SPM, converting the data stream into the semantically segmented image of domain walls and the corresponding uncertainty. We further demonstrate the pre-selected experimental workflows on thus discovered domain walls, and report alternating high- and low-polarization dynamic (out-of-plane) ferroelastic domain walls in a (PbTiO$_3$) PTO thin film and high polarization dynamic (out-of-plane) at short ferroelastic walls (compared with long ferroelastic walls) in a lead zirconate titanate (PZT) thin film. This work establishes the framework for real-time DCNN analysis of data streams in scanning probe and other microscopies and highlights the role of out-of-distribution effects and strategies to ameliorate them in real time analytics.

[a] sergei2@utk.edu
[b] ziatdinovma@ornl.gov

Ferroelectric materials are one of the most exciting materials systems explored over the last century. Traditionally, these materials have been explored in the context of bulk applications, including piezoelectric transducers and actuators and electrooptical devices.[1] The progress in sol-gel and pulsed laser deposition methods driven by the development of microelectromechanical systems[2] and ferroelectric-based information technology[3, 4] devices have stimulated research towards the ferroelectric thin films and nanosystems.[2, 5] Finally, recent advanced in materials characterization have opened for exploration the device concepts enabled by the functionality of individual topological elements such as domain walls[6] and vortices.[7]

Over the last decade, the field of ferroelectric materials has been significantly broadened by the discovery of ferroelectric phenomena in new materials classes, most notably binary oxides such as hafnia and zirconia and their solid solutions;[8-12] scandium, boron, and aluminum nitrides;[13-17] and zinc-magnesium oxides.[18, 19] These discoveries have both changed the extant paradigms on mechanisms underpinning ferroelectricity, and open the pathway to a broad range of information technology device applications. Similarly, ferroelectric phenomena were discovered in low-dimensional materials such as twisted double layers graphene,[20] and boron nitride,[21] molybdenum disulfide,[22] tungsten ditelluride,[23] etc.

Both in classical ferroelectric materials and new low dimensional materials systems, the phenomena of interest, including polarization switching, conductivity and conductance switching, etc. are strongly coupled to the structural and topological defects.[6, 24, 25] In classical ferroelectric, polarization dynamics, including domain nucleation and 180° ferroelectric wall pinning is strongly affected by the presence of the ferroelastic domain walls or structural defects.[26] In low dimensional ferroelectrics, the structural defects, stacking ordering, and heterostructure are coupled to the polarization dynamics.[21, 23, 27] Similarly, coupling between polarization discontinuities and electronic systems gives rise to a broad set of wall-coupled transport phenomena.[28-31]

Many of these advances in the physics of low-dimensional ferroelectrics have been enabled by scanning probe microscopy (SPM) techniques, most notably Piezoresponse Force Microscopy (PFM)[32-34] and conductive Atomic Force Microscopy (cAFM). In these measurements, the sharp SPM tip serves as a mobile electrode and sensor, applying a bias to the material's surface and detecting associated electromechanical responses and currents. In the imaging modes, these responses are collected in the form of dense 2D maps. However, spectroscopic applications such as hysteresis loop measurements or current-voltage curve mapping required the development of hyperspectral imaging modes, collecting the spectra over a dense rectangular grid.[35, 36] The subsequent physics-based or unsupervised machine learning-based analysis of the resultant multidimensional data sets yields the 2D images highlighting the ferroelectric or conductivity functionality of individual microstructural elements.[37-39] Similarly, unsupervised ML methods can be used to build structure-property relationships in ferroelectric materials.[40]

However, exploration of domain wall physics via the combination of the PFM and spectroscopic images suffers from a number of significant limitations. First, the typical size of domain structures in materials is fairly large (> 100 nm), whereas the thickness of domain walls is typically small.[1, 41, 42] Consequently, the position of the probing tip with respect to the wall can vary strongly, limiting quantitative studies. Secondly, the polarization distributions in the material can change during the measurements, e.g. polarization switching can shift both 180º and ferroelastic walls.[43-46] In this case, establishing the relationship between the domain and hyperspectral image is complicated.

Here we explore the intrinsic functionality of the ferroelastic domain walls via an automated experiment in PFM. We develop a workflow for real-time data analytics based on the ensembled ResHedNet architecture that minimizes out-of-distribution drift effects and yields domain wall structure and reconstruction uncertainty. We further extend this approach to investigate the properties at domain walls, i.e., perform band excitation piezoresponse spectroscopy measurement at domain walls to study the properties encoded in piezoresponse-voltage hysteresis loops. We observed that adjacent ferroelastic domain walls exhibit alternating strong and weak out-of-plane polarization, attributed to the wall tilting effects. However, we also observe more subtle variations in domain wall switching behavior associated with proximity to large in-plane domains.

As a model system, we have chosen a $PbTiO_3$ (PTO) thin film grown by metalorganic chemical vapor deposition method on a $SrRuO_3$ conducting layer and (001) $KTaO_3$ substrates.[47] Previously, we have explored the relationship between local domain structures and the ferroelectric properties encoded in the piezoresponse-voltage hysteresis in this sample implementing an active learning network on operational piezoresponse force microscopy. That work demonstrated different exploration pathways (sequence of sampled points) by the active learning workflow for the acquisition functions based on on-field and off-field hysteresis loops, suggesting different dominating properties of the hysteresis loops under on-field and off-field conditions.[48] In addition, our analysis of causal physical mechanism among different band excitation PFM channels revealed a mutual interaction among the channels associated with different properties.[49] These previous studies aimed at discovering microstructural elements that maximize certain functional response of interest, i.e. find the domain structure (or other microstructural elements) that corresponds to the largest hysteresis loop opening.

Here, we demonstrate an alternative approach for an automated experiment based on the identification of a priori known object of interest (i.e. ferroelastic walls) in real time and further exploring its functionality, implemented with ensembled ResHedNet on operational PFM.

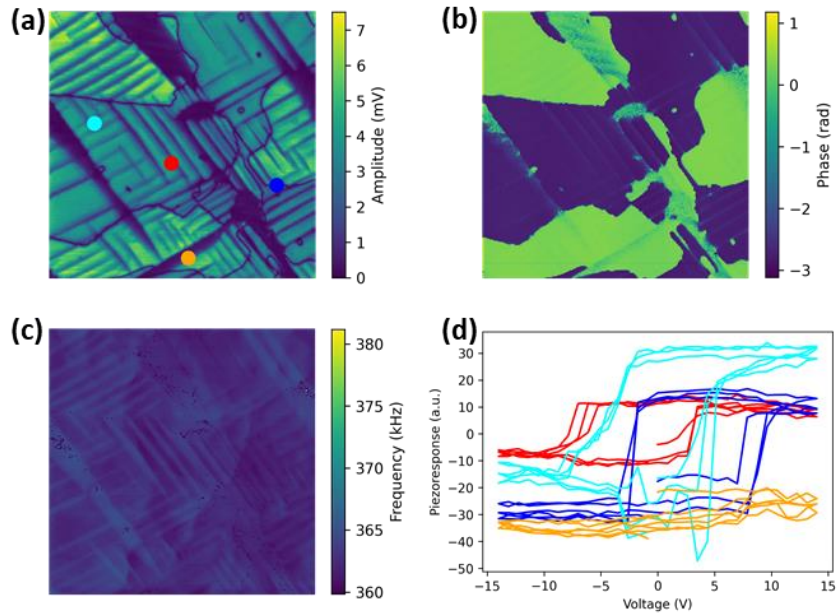

**Figure 1.** Typical domain structure and piezoresponse-voltage loops in PTO sample. **(a)-(c)** band excitation piezoresponse force microscopy amplitude, phase, and resonance frequency images, respectively. **(d)** piezoresponse-voltage loops at locations marked on **(a)**. Note the complexity of the ferroelectric domain patterns in **(a)** and broad variability of the hysteresis loops across the surface in **(d)**.

The typical domain structure of the PTO film is shown in Figure 1. Here, the large dark domains in amplitude image (Figure 1a) corresponding to the regions with the in-plane polarization are clearly seen. At the same time, the alternating bright out-of-plane c-domains and dark in-plane *a*-domains show as clearly visible stripes (Figure 1a), corresponding to *a-c* domain structures. In addition, dark meandering lines are the $180^0$ domain walls between antiparallel $c^+$ and $c^-$ domains, $c^+/c^-$ domains are visible in phase image (Figure 1b). Note that several interesting domain phenomena can be directly seen from the image, including the domain splitting in the vicinity of the large *a*-domains to accommodate the stress. Resonance frequency image also illustrates clear ferroelastic domains (Figure 1c); here the crystallographic orientation variation gives rise to different strain and elastic properties between *a*-domains and *c*-domains, and leads to the domain contrast in resonance frequency. Figure 1d shows example piezoresponse-voltage loops from several representative locations, including *a*-domain (orange), *c*-domain (cyan), $c^+/c^-$ 180º domain wall (blue), and *a-c* domain wall (red). The difference between these loops suggests the roles of location domain/wall structure in material's functionality.

Here, we aim to explore the physics of specific domain walls, namely explore the polarization switching at selected types of ferroelastic walls. These can be determined on the image directly via human eye; however, finding these automatically is a challenge. Over the last 5 years, deep convolutional neural networks have been broadly adopted in electron[50, 51] and scanning probe microscopies.[52-54] However, while these techniques have amply demonstrated their potential for post-acquisition data analysis, their implementation as a part of real time experiment is highly non-trivial. The reason for this is the out of distribution drift effects, manifested as network trained for one set of imaging conditions will perform poorly when the image conditions change. These out of distribution effects have been broadly recognized in the ML community, and are directly responsible for dearth of practical applications of ML methods in medicine or autonomous driving.[55-62]

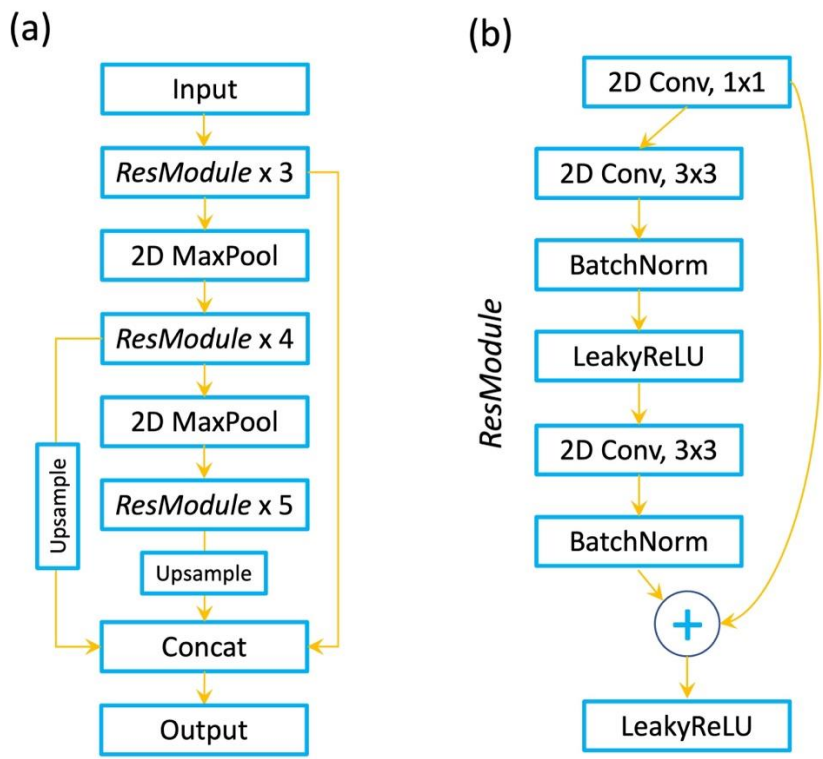

**Figure 2.** (a) Schematic of the ResHedNet model. It consists of three residual blocks with 3, 4, and 5 residual modules. The schematic of an individual residual module is shown in (b). The outputs of different blocks are fused to produce the final prediction. Note that the input to second (third) block is the max-pooled output of the first (second) block. See the accompanying notebook for more details.

We chose a deep learning-based semantic segmentation for converting raw microscopy data into the domain walls maps. Earlier, we found that standard neural network architectures for semantic segmentation such as U-Net[63] do not work well on thin objects such as domain walls. We demonstrated that augmenting the modified version of holistically nested edge detector[64] with ResNet[65] modules leads to significant improvement (+30%) in the domain wall detection.[26] However, the previous work was limited to the already collected data where training and test/validation data came from the same distribution (i.e., the same experiment). Application of the network trained on data from previous experiments to a new experiment under slightly different conditions often leads to the unreliable performance, the behavior generally referred to as the out-of-distribution shift. Very generally, it can be understood as the same object visualized under slightly different conditions will be perceived by the neural network differently, and once imaging conditions are sufficiently different the identification becomes impossible. While this problem has been recognized only few years ago (and is implicitly responsible for the lack of the fully autonomous driving to date), it is crucial for any real-time applications of ML methods during the experiments.

The two most important requirements for working with data characterized by a distribution shift are robust generalization and reliable uncertainty estimates. The fully Bayesian approach to deep learning based on Hamiltonian Monte Carlo sampling[66, 67] - where neural network weights are substituted by probability distributions - is considered to be the current state-of-the-art for these purposes. In this approach, the predictive mean and uncertainty on the newly observed data $x_*$ given the training set $D$ are expressed as

$$P(x_*|D) = \int_\theta P(x_*|w)P(w|D)dw \approx \frac{1}{N}\sum_{n=1}^{N} P(x_*|w^n, D) = \hat{f}_*, \tag{1a}$$

$$U[f_*] = \frac{1}{N}\sum_{n=1}^{N}(f_*^n - \hat{f}_*)^2, \tag{1b}$$

where $w^n \sim P(w|D)$ are neural network weights drawn from the posterior. However, the fully Bayesian approach to deep learning is computationally expensive and is not suitable for real-time implementation even on modern GPU accelerators.

A recent careful empirical review of many of the available approximation techniques to the fully Bayesian deep learning suggests that the deep ensembles approach could be the most promising and easy-to-implement solution.[68, 69] In the deep ensembles method, one trains a collection of neural networks via a stochastic gradient descent using different random initialization of weights and different shuffling of training samples. The deep ensembles can be viewed as approximating the posterior predictive distribution with the set of point masses at different modes such that the posterior samples $w^n$ in Eq. (1) are replaced with different ensemble weights. We note that deep ensembles will still fail on data characterized by very large distribution shifts due to inherent correlative nature of deep learning techniques. Hence, during the ensemble training, we also perform data augmentation on-the-fly to account for a likely variation in imaging conditions and domain wall orientations to minimize the distribution shift between training and application domains. To further improve the prediction capacity of the deep ensemble, we perform a stochastic weight averaging[70] at the end of the training trajectory of each individual model.

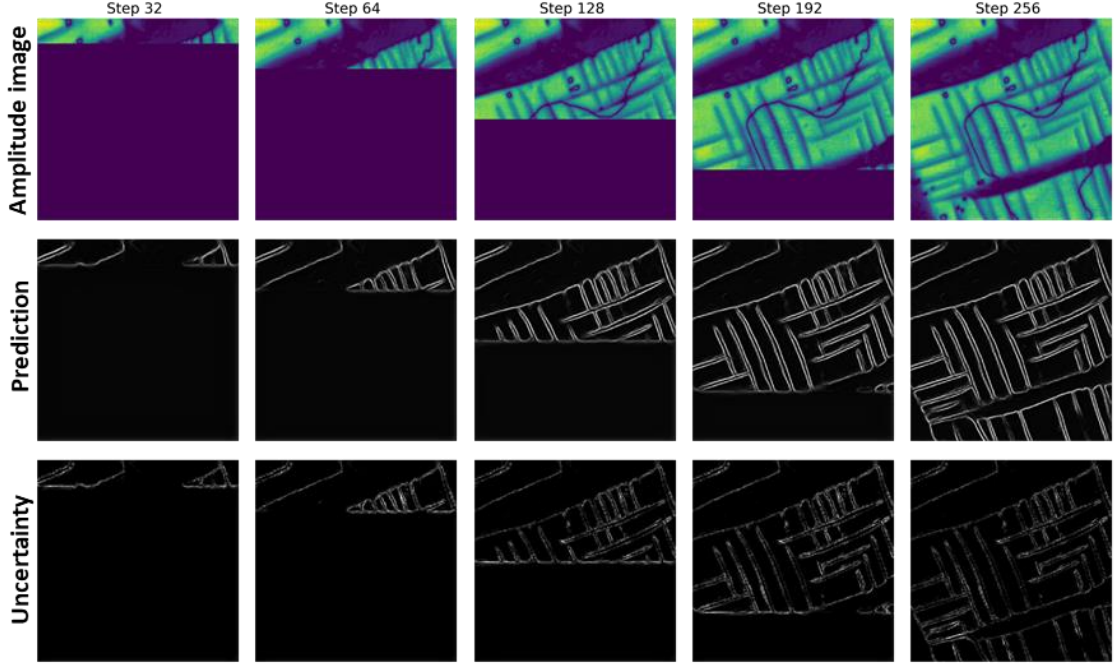

**Figure 3.** Simulated experiment process with pre-acquired data. A 256*256 BEPFM amplitude image is added line-by-line as the raw data for ResHedNet prediction; here ResHedNet mean prediction and uncertainty from several representative steps are shown. The uncertainty at the domain walls likely originate from a different contrast gradient at domain walls. Note also a high uncertainty at the current scan line on all images.

The ensembled ResHedNet model makes generally good predictions of ferroelastic domain walls from BEPFM amplitude image. Before implementing this ResHedNet model to operating microscopes, we first simulated the experiment process with pre-acquired data, as shown in Figure 3. In the simulated experiment, a BEPFM amplitude data is added as the raw data for ResHedNet model analysis line-by-line, analogous to the SPM scanning process where data is acquired line-by-line. ResHedNet predictions and uncertainties from several representative steps are shown in Figure 3, indicating a good performance of the ensembled ResHedNet model in analyzing on-the-fly data. The uncertainty levels are comparable to those on the 'hold-off' set at the model training stage, indicating a good generalization.

Next, we implement the trained ResHedNet model in operating microscope to realize real-time prediction of ferroelastic domain walls. To enable parallel SPM scan and ResHedNet analysis, we developed and deployed a workflow shown in Figure 4a and integrated multiple software (LabView, Jupyter Notebook, and Igor) and Hardware (Asylum Research Cypher microscope, Field Programmable Gate Arrays (FPGA), National Instruments DAQ card). To enable the programmatic control of the whole system, the FPGA receives orders from Jupyter Notebook and digitally communicates with Cypher microscope and National Instruments DAQ card. In a BEPFM measurement, Jupyter Notebook sends orders of performing line scan and performing ResHedNet analysis alternatingly as shown in Figure 4a. The real-time BEPFM data and ferroelastic domain wall image are visible in the Jupyter Notebook.

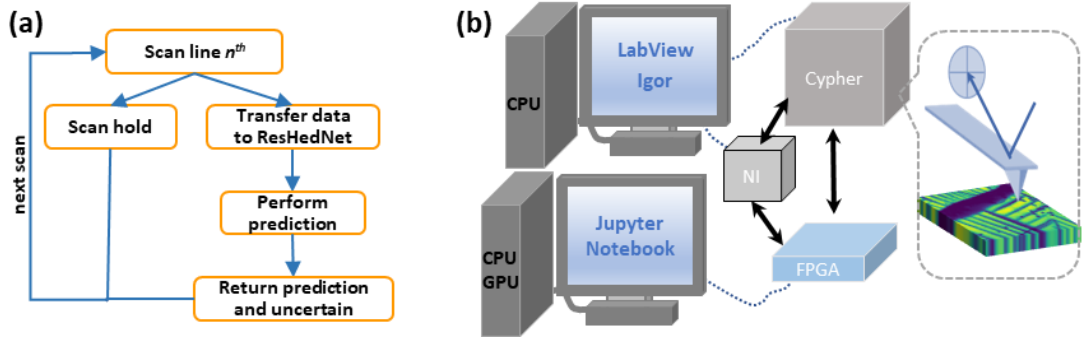

**Figure 4.** The workflow and integrated system for the real-time ferroelastic wall investigation.

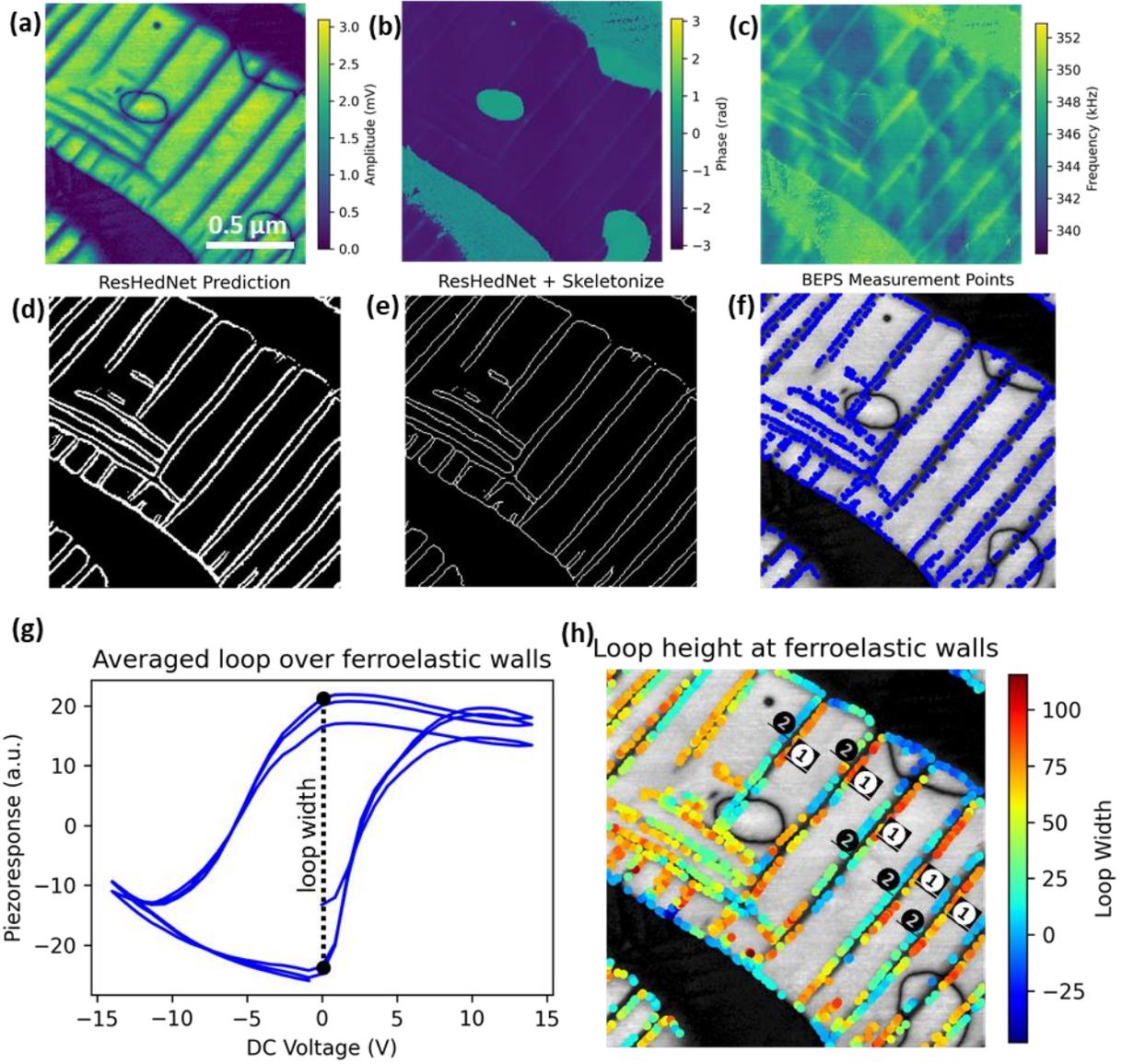

**Figure 5.** ResHedNet ensemble on operating PFM to explore ferroelastic wall polarization dynamics in PTO. (a-c), BEPFM amplitude, phase, and resonance frequency images of the PTO film; (d) ResHedNet prediction on ferroelastic domain walls; (e) Skeleton of ResHedNet prediction; (f) BEPS measurement points on ferroelastic domain walls; (g) averaged piezoresponse vs. voltage loops over all measurement points, and schematically shows how the loop width is extracted; (h) loop width over the ferroelastic domain walls, where the color represents loop width.

Shown in Figure 5 is the results of the model system PTO obtained by this approach. The predictive uncertainty was monitored throughout the experiment to ensure the new data is within the training data distribution. Figure 5a-c show the BEPFM amplitude, phase, and resonance frequency images. The trained ResHedNet predicts the ferroelastic domain wall image (Figure 5d) from the amplitude image (Figure 5a). Then, a scikit-image[71] 'skeletonize' function was applied to generate the wall skeleton image (Figure 5e). According to the wall skeleton image, 1/3 of wall locations were evenly selected as the band excitation piezoresponse spectroscopy (BEPS) measurement points, as shown in Figure 5f. Figure 5g shows the averaged BEPS loops from these points at ferroelastic domain walls. In order to visualize the relationship between wall structure and BEPS loops, the loop width (as indicated in Figure 5g) at each point was extracted as a descriptor of out-of-plane polarization dynamics and plotted in Figure 5h. Interestingly, it shows that high-loop-width walls (labeled as wall ①) and low-loop-width walls (labeled as wall ❷) are alternatingly distributed. We ascribe this behavior to the tilting of ferroelastic domain walls as has originally been explored by Ganpule for polarization switching.[72-75] Here, switching of the wall oriented away from the tip and towards the tip involves dissimilar number of in-plane domains, leading to dissimilar responses.[46]

The analyses of hysteresis loops are also extended to coercive field, loop area, and nucleation biases, as show in Figure 6. Loops from representative locations (e.g., large loop area and small loop area locations) are also shown.

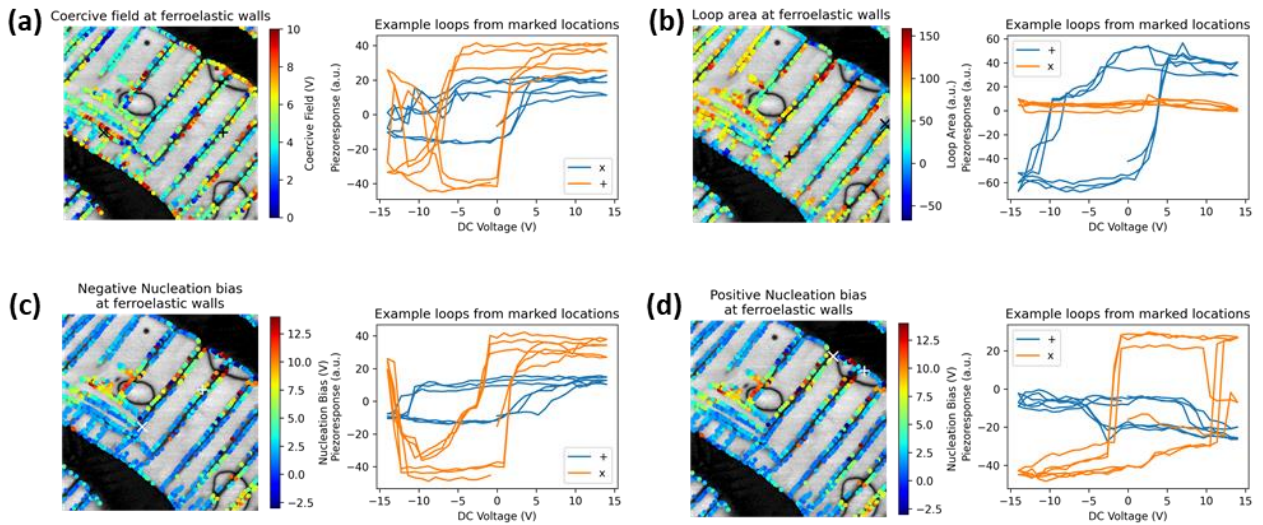

**Figure 6.** Analyses of coercive field, hysteresis loop area, and nucleation bias, and representative loops from marked locations. (a)-(d), Coercive field, hysteresis loop area, negative nucleation bias, positive nucleation bias distribution at ferroelastic domain walls, and representative loops from the corresponding marked locations.

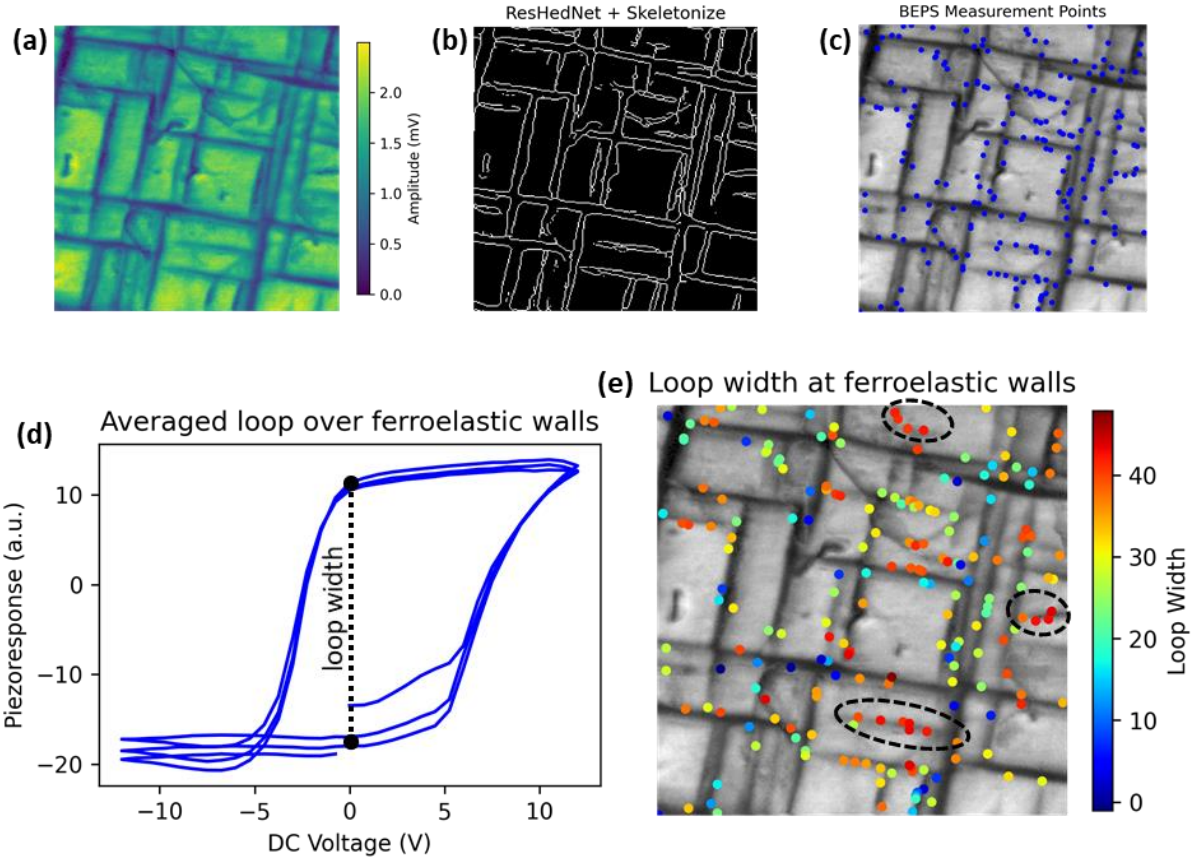

**Figure 7.** ResHedNet ensemble on operating PFM to explore ferroelastic wall polarization dynamics in PZT. (a), BEPFM amplitude image; (b) Skeleton of ResHedNet prediction on ferroelastic domain walls; (c) BEPS measurement points on ferroelastic domain walls; (d) averaged piezoresponse vs. voltage loops over all measurement points, and schematically shows how the loop width is extracted; (e) loop width over the ferroelastic domain walls, where the color represents loop width.

This approach is further used to study a PZT thin film, results shown in Figure 7. The BEPFM amplitude shows (Figure 7a) both long ferroelastic domains over the whole image and short needle domains with terminations. ResHedNet prediction (Figure 7b) shows both long and short domain walls, the BEPS measurement was performed at 1/250 of domain wall points here (Figure 7c). Figure 7d shows the averaged piezoresponse vs. voltage loop at these points and Figure 7e shows the loop width at these points. In this case, it is observed that the short walls with terminations generally show high-loop-width, as indicated by dashed circles. This is most likely because that the bulk polarization is out-of-plane under these in-plane short *a*-domains.

To summarize, we developed and deployed on an operational microscope a workflow directly combining the DCNN analysis of a prior known objects of interest and real time spectroscopic measurements. This required performing DCNN analysis on real-time data generation process in the presence of the out of distribution drift effects. To solve this problem, we developed ensembled ResHedNet that minimizes the out of distribution drift effects. The ensembled ResHedNet was implemented on operating PFM to perform real time investigation of ferroelastic domain walls. Using this approach, we studied piezoresponse dynamics at ferroelastic domain walls of two thin film materials, PTO and PZT. The PTO sample shows alternatingly distributed high- and low- piezoresponse dynamics ferroelastic domain walls, and the PZT sample shows high piezoresponse dynamics at the short ferroelastic walls compared to long ferroelastic walls. This approach is universally applicable for real-time data analytics in other imaging techniques, such as transmission electron microscopes, secondary ion mass spectrometry, and optical microscopes.

**Methods:**

*Preparation of training and validation dataset:*

To prepare the training and validation datasets for the model training phase, the ferroelastic domain walls in BEPFM amplitude images were labeled using ImageJ to generate the ground truth a-c wall structure images. Then, a data set of paired raw images and ground truth wall images was created using AtomAI[76] to extract sub-images from BEPFM amplitude and a-c wall structure images. Finally, this data set was randomly split into training dataset and validation dataset. In this work, our training + validation data set contained 18000 pairs of 250*250 pixels raw and ground truth images created from nine source images, where 80% was training data and 20% was validation data. The AtomAI trainers were used to perform model training. The ensembled ResHedNet network was trained for 2000 iterations with the batch size of 10 and consisted of 10 models. More details about the use of ensembled ResHedNet model and training process can be found in the provided Jupyter Notebook.

**Acknowledgements:**
This effort (implementation in SPM, measurement, data analysis) was primarily supported by the center for 3D Ferroelectric Microelectronics (3DFeM), an Energy Frontier Research Center funded by the U.S. Department of Energy (DOE), Office of Science, Basic Energy Sciences under Award Number DE-SC0021118. This research (ensemble-ResHedNet) was sponsored by the INTERSECT Initiative as part of the Laboratory Directed Research and Development Program of Oak Ridge National Laboratory, managed by UT-Battelle, LLC for the US Department of Energy under contract DE-AC05-00OR22725. The research was partially supported (piezoresponse force microscopy) at Oak Ridge National Laboratory's Center for Nanophase Materials Sciences (CNMS), a U.S. Department of Energy, Office of Science User Facility.

**Conflict of Interest**

The authors declare no conflict of interest.

**Authors Contribution**

S.V.K. and M.Z. conceived the project. M.Z. realized the ensemble-ResHedNet. Y.L. performed analyses. Y.L. deployed the ResHedNet to PFM measurement and obtained results. K.K. helped with the deployment. H.F. provided the PTO and PZT sample. All authors contributed to discussions and the final manuscript.

**Data Availability Statement**

The data that support the findings of this study are available at https://github.com/yongtaoliu/Ensemble-ResHedNet.